\documentclass[%
 aip,
rsi,%
 amsmath,amssymb,
 reprint,%
]{revtex4-1}

\usepackage{graphicx}
\usepackage{dcolumn}
\usepackage{bm}
\usepackage{hyperref}
\usepackage{tikz}
\usetikzlibrary{shapes,arrows}
\usetikzlibrary{positioning}
\tikzstyle{decision} = [rectangle, draw, fill=red!20, 
    text width=8em, text centered, rounded corners, minimum height=2em]
\tikzstyle{block} = [rectangle, draw, fill=blue!20, 
    text width=7em, text centered, rounded corners, minimum height=2em]
\tikzstyle{blockp} = [rectangle,dashed, draw, fill=blue!20, 
    text width=4.5em, text centered, rounded corners, minimum height=2em]
    \tikzstyle{blockt} = [rectangle, draw, fill=gray!20, 
    text width=5.5em, text centered, rounded corners, minimum height=2em]
\tikzstyle{line} = [draw, -latex']
\tikzstyle{cloud} = [draw, ellipse,fill=green!20, minimum height=2em]

\newcolumntype{d}[1]{D{.}{.}{#1}}
\usepackage{amsmath}
\DeclareMathOperator{\Tr}{Tr}

\begin{document}


\title{Calculation of the Anisotropic Coefficients of Thermal Expansion: A First-Principles Approach}

\author{Nicholas A. Pike}
\email{Nicholas.pike@smn.uio.no}
\affiliation{Centre for Materials Science and Nanotechnology, University of Oslo, NO-0349 Oslo, Norway}
\author{Ole M. L{\o}vvik}%
\affiliation{Centre for Materials Science and Nanotechnology, University of Oslo, NO-0349 Oslo, Norway}
\affiliation{SINTEF Materials Physics, Forskningsveien 1, NO-0314 Oslo, Norway}

\date{\today}

\begin{abstract}
Predictions of the anisotropic coefficients of thermal expansion are needed to not only compare to experimental measurement, but also as input for macroscopic modeling of devices which operate over a large temperature range. While most current methods are limited to isotropic systems within the quasiharmonic approximation, our method uses first-principles calculations and includes anharmonic effects to determine the temperature-dependent properties of materials. These include the lattice parameters, anisotropic coefficients of thermal expansion, isothermal bulk modulus, and specific heat at constant pressure. Our method has been tested on two compounds (Cu and AlN) and predicts thermal properties which compare favorably to experimental measurement over a wide temperature range.
\end{abstract}


\maketitle


\section{Introduction}
While experimental measurements of the coefficient of thermal expansion (CTE) can be done using a number of experimental techniques~\cite{Belousov2007, Cliffe2012, Jones2013, Samovilov2006, Dudescu2006}, most theoretical tools available for predicting the CTE are limited to the quasiharmonic approximation (QHA) or the  Debye-Gruneisen approximation. There are a few programs~\cite{Togo2015, Togo2010} and several models~\cite{Jin2002, Quong1997, Tohei2016, Liu2017, Wang2006, Wang2011, Ding2015, Lamowski2011, Liu2018a,Gilev2018} available for scientists to calculate the CTE, and its associated thermal properties, from first-principles calculations. These currently available methods principally rely on the QHA. They allow the user to determine the temperature dependence of the isotropic lattice parameters, the coefficient of volume expansion, isothermal bulk modulus and specific heat at constant pressure. Since these programs determine the coefficient of volume expansion, they are limited to calculations of systems with a considerable amount of structural symmetry. A limited amount of work has been published on the thermal properties of anisotropic systems employing anharmonic effects~\cite{Shulumba2016,Shulumba2015,Schmerler2014}, but this work is not available as a comprehensive package of tools that can automatically predict such properties from the crystal structure of a material.

In this work, an algorithm is developed to determine the thermal properties of both isotropic and anisotropic systems using density functional theory (DFT) and density functional perturbation theory (DFPT) calculations. This algorithm utilizes the Vienna Ab initio Simulation Package (\textsc{vasp})~\cite{Kresse1993, Kresse1996a, Kresse1996b} for DFT and DFPT calculations and the Temperature Dependent Effective Potential (\textsc{tdep})~\cite{Hellman2013, Hellman2013b, Hellman2011} software package to calculate the ground state and interatomic force constants using a method designed to work for both harmonic and strongly anharmonic systems~\cite{Hellman2011}. 

To demonstrate the abilities of our methods, two systems with distinctive symmetries are compared to a variety of experimental results. The first system, pure copper (Cu)~\cite{Hahn1970, White1973}, possesses cubic symmetry and therefore the CTE is a single temperature-dependent value. The temperature dependent properties of copper are thoroughly investigated experimentally~\cite{Touloukian1975, Hahn1970, Adenstedt1936, Carr1964, Carr1964a, Esser1941, Fraser1965, Lifanov1968, Rubin1954, Arblaster2015} and Cu is consequently used to determine the quality of our calculation methods. The second material, aluminum nitride (AlN)~\cite{Figge2009}, has a hexagonal crystal structure and therefore two unique non-zero CTE's and is used to demonstrate our ability to calculate the expansion coefficients of an anisotropic material with a known temperature dependent anharmonic interaction~\cite{Shulumba2016}.

In section~\ref{calculation_methods}, an outline of the calculation methods is provided and in section~\ref{sec:2}, we analyze the results of our methods via comparison to available experimental data for Cu and AlN. Finally, in section~\ref{sec:3}, we provide a concluding discussion. An appendix is provided that outlines the algorithm used to generate the necessary files for the \textsc{vasp} and \textsc{tdep} calculations, gather the necessary results, and perform the needed temperature dependent calculations.

\section{Calculation Methods}\label{calculation_methods}
Our software launches both DFT and DFPT calculations using \textsc{vasp}~\cite{Kresse1996a, Kresse1996b}. The present benchmark calculations were performed with the projector-augmented wave (PAW) method~\cite{Blochl1994} to describe the core electrons by utilizing PAW pseudopotentials~\cite{Kresse1999}. Exchange and correlations effects were described by the Perdew-Burke-Ernzerhof (PBE) generalized gradient approximation (GGA)~\cite{Perdew1996}. To describe the electronic system, a plane-wave energy cutoff equal to 500~eV was used and a Monkhorst-Pack~\cite{Monkhorst1976} mesh of points was generated for each grid in reciprocal space assuming a $k$-point density of at least five points per \AA$^{-1}$. For the relaxation, ground-state, and configuration calculations the same $k$-point mesh was used.  DFPT calculations were done on a $k$-point mesh twice as dense as the grid used in the ground state calculations.

For each ground state and configuration calculation, the iterations of the total energy were stopped once the differences in energy between successive iterations were less than 0.01 meV. To calculate the dielectric, elastic, and Born effective charge of our material, we used finite-differences as implemented within \textsc{vasp} where only symmetry-inequivalent atomic displacements were used to calculate the Hessian matrix. The elastic tensor was used to determine the Debye temperature and the isentropic bulk modulus, as outlined below. The dielectric tensor and Born effective charge tensor were used to account for the long-range electrostatic interactions in semiconducting compounds within \textsc{tdep}.

It is important to note that there is a fundamental relationship between the isoentropic bulk modulus (via the elastic tensor) and the isothermal bulk modulus (found by fitting the equation of state) directly related to the ratio of specific heats. While this ratio is nearly identical to one (see below) we will distinguish here due to the differences in the methodology used to determine the quantities.

\begin{figure}[t!]
\includegraphics[width=0.45\textwidth]{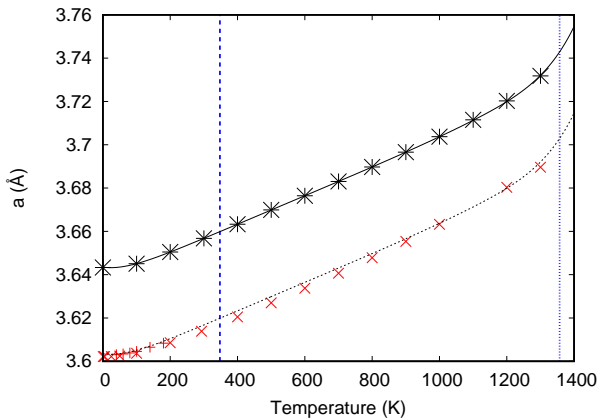}
\caption{\label{thermal_lattice_copper}Calculated and experimental temperature dependent lattice parameters for Cu. Our theoretical calculations are shown as black points (every 50th point is shown) and the solid line corresponds to an eighth order polynomial fit. The black dashed line equals the solid line except for a constant shift correcting for the normal error seen in PBE GGA calculations of the ground state lattice constant. Experimental data for the lattice parameters in red from Ref.~\citenum{Shah1944} (plus signs) and the recommended lattice parameter of Ref.~\citenum{Touloukian1975} are shown as X's. The blue dashed and dotted vertical lines denote the Debye and melting temperature respectively. }
\end{figure}

Using the elastic tensor, one can determine both the Debye temperature and isentropic bulk modulus of the system using the following relationships. First, the Voigt average of the bulk ($B$) and shear ($G$) moduli (upper bound for polycrystalline materials) are found from the components of the elastic tensor, $C_{ij}$ ($i,j = 1..6$), as~\cite{Nye1957,bulk_pike}
\begin{align}
    B &= \frac{1}{9}\bigg((C_{11} +C_{22}+C_{33})+2(C_{12}+C_{31}+C_{23})\bigg) \nonumber \\
    G &= \frac{1}{15}\bigg( (C_{11} +C_{22}+ C_{33}) -(C_{12}+C_{31}+C_{23})\nonumber \\ &+3(C_{44}+C_{55}+C_{66})\bigg).
\end{align}

\begin{figure}[t!]
\includegraphics[width=0.45\textwidth]{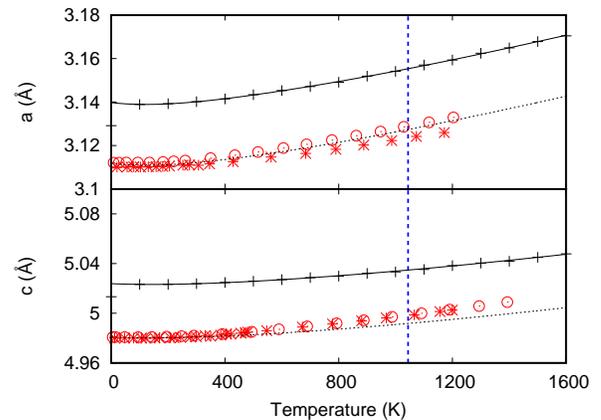}
\caption{\label{thermal_lattice_aln} Calculated and experimental temperature dependent lattice parameters for AlN. Our theoretical calculations are shown as black points (every 50th point is shown) and the solid line corresponds to an eighth order polynomial fit. The dashed black lines are equal to the solid lines except for a constant shift correcting for the normal error seen in PBE GGA calculations of the ground state lattice constants. Experimental data for the lattice parameters from Refs.~\citenum{Figge2009} (asterisk) and~\citenum{Wang2011} (open circles) are shown. The blue dashed vertical line denotes the Debye temperature. }
\end{figure}

With the bulk and shear moduli one can then calculate the longitudinal and shear sound velocities ($v_l$ and $v_s$) as~\cite{Jia2017}
\begin{equation}
    v_l =\sqrt{\frac{B+4/3G}{\rho}}
\end{equation}
and
\begin{equation}
    v_s =\sqrt{\frac{G}{\rho}},
\end{equation}
where $\rho$ is the density of the material. Then, the Debye temperature can be calculated as~\cite{Jia2017}
\begin{equation}
    \Theta_D = \frac{h}{k_B}\left[\frac{9m}{4\pi}\right]^{1/3}\left[\left(\frac{1}{v_l^3}+\frac{2}{v_s^3}\right)\right]^{-1/3},
\end{equation}
where $h$ is Plank's constant, $k_B$ is Boltzmann's constant, and $m$ is the number of atoms per volume and, unlike in Ref.~\citenum{Jia2017}, we do not separate the acoustic bands and use all the phonon bands to calculate the Debye temperature.
\begin{figure}[t!]
\includegraphics[width=0.4\textwidth]{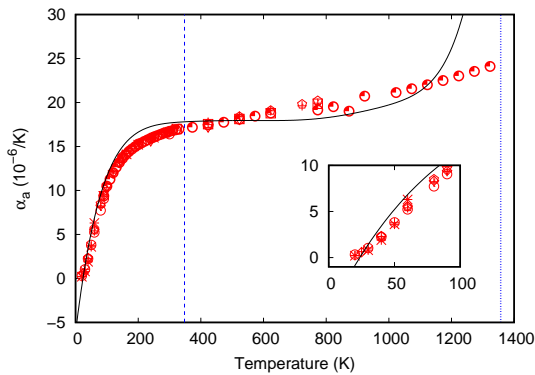}
\caption{\label{thermal_expansion_copper} The CTE for Cu from our theoretical calculations is drawn with a solid line and experimental data as red symbols: plus signs (Ref.~\citenum{Hahn1970}), crosses (Ref.~\citenum{Adenstedt1936}), asterisks (Ref.\citenum{Carr1964}), open squares (Ref.~\citenum{Esser1941}), open circles (Ref.~\citenum{Fraser1965}), open inverted triangles (Ref.~\citenum{Lifanov1968}), open diamonds (Ref.~\citenum{Rubin1954}), open pentagons (Ref.~\citenum{Simmons1963}) and quarter open circles (Ref.~\citenum{Leksina1963}). Inset: Calculated and experimental data points for temperatures between 0 and 100 K indicating that our calculations are only accurate for $T > 30$ K. The blue dashed and dotted vertical lines denote the Debye and melting temperature respectively. }
\end{figure}

Since the Debye temperature is calculated from the elastic tensor, which is calculated at zero temperature within~\textsc{vasp}, it also represents the ground state (0~K) Debye temperature. While one could, in principle, calculate the elastic properties as a function of temperature~\cite{Steneteg2013}, and therefore the Debye temperature as a function of temperature, we have found that the Debye temperature at zero temperature provides an adequate starting point for our calculations since variations in the configuration temperature, outlined below, do not significantly modify our results.

The Debye temperature is used within \textsc{tdep} to generate an initial guess for the force constants of the system. They are in turn employed to generate configurations based on a canonical ensemble.  Here, twelve configurations are generated as follows: with the Debye temperature and symmetry of the cell, one generates an initial guess for the interatomic force constants and solves the equations of motion for the system by finding the resulting eigenvalues and eigenvectors.  Then, these eigenvalues and eigenvectors are used to determine the amplitude and velocity of each atom chosen such that these quantities are normally distributed over a canonical ensemble. The configurations consisted in this study of 208 atoms for Cu and AlN. The ensemble is generated at a finite temperature (here designated the configuration temperature) with Bose-Einstein statistics used to determine the mean normal mode amplitude ~\cite{West2006}. \textsc{Vasp} is then used to calculate the energies, displacements, and forces for each of the generated configurations. These are combined within \textsc{tdep} to generate the finite-temperature force constants by fitting the Born-Oppenheimer energy surface. This procedure allows us to go beyond the quasiharmonic approximation by explicitly including anharmonic effects through the canonical ensembles.  
\begin{figure}[t!]
\includegraphics[width=0.4\textwidth]{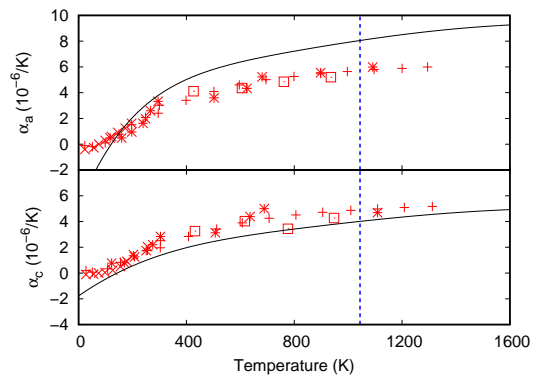}
\caption{\label{thermal_expansion_aln}The calculated and experimental CTE for AlN with our theoretical calculations correspond to solid lines and experimental data for the thermal expansion coefficients as red symbols: plus signs (Ref.~\citenum{Wang2011}), crosses (Ref.~\citenum{Ivanov1997}), asterisks (Ref.~\citenum{Slack1975}), and open squares (Ref.~\citenum{Yim1974}). The blue dashed vertical line denotes the Debye temperature.}
\end{figure}

The extracted force constants are used to find the phonon frequencies and phonon density of states on a $q$ mesh  in reciprocal space (in this work $30\times 30\times 30$ points were used) using the tetrahedron integration approach~\cite{Lehmann1972}. The density of states is used to calculate thermal properties, as outlined below. Using this grid and integration method ensured the convergence of the free energy to within 0.01 meV/atom.

The calculated lattice parameters ${\bf x}(T,p)$ depend on temperature and pressure. Determining the lattice parameters at ambient pressure ${\bf x}(T,0)$ requires minimizing the free energy $F({\bf x},T)$ at each temperature. $F$ can be expressed as a function of the total electronic energy $U(\bf{x})$ and the vibrational free energy ($F_{\textrm{vib}}({\bf x},T)$) as
\begin{align}
\label{free_energy}
F({\bf x},T) &= U({\bf x}) + F_{\textrm{vib}}({\bf x},T), \nonumber \\
F_{\textrm{vib}}({\bf x},T) &= \int^{\omega_{max}}_0 \sum_\lambda \Bigg[g(\omega_\lambda)\frac{\hbar\omega_\lambda}{2}  \nonumber \\  &+k_BTg(\omega_\lambda) \ln \left( 1-\exp\left(\frac{\hbar\omega_\lambda}{k_B T}\right)\right)d\omega_\lambda\Bigg],
\end{align}
where $\lambda$ is the phonon mode index and both the vibrational density of states ($g(\omega)$) and phonon frequencies $\omega_\lambda$ depend on $\bf{x}$. $U$ is the volume dependent total electronic energy of the system. In order to minimize the free energy, a set of lattice parameters is generated automatically by applying strain to the system around the ground state equilibrium lattice parameters. This was in the present work done by finding a series of six lattice parameters in each symmetry unique direction such that the new lattice parameters range between one percent compressive and four percent tensile strain. This provided a smooth free energy surface at each temperature.  
\begin{figure}[t!]
\includegraphics[width=0.4\textwidth]{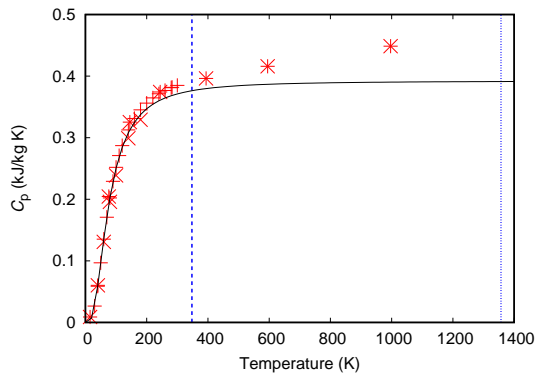}
\caption{\label{comp_cv_copper} Calculated specific heat at constant pressure for Cu. Literature data for $C_p$ is given as red points: plus signs from Ref.~\citenum{White1973}, crosses from Ref.~\citenum{Shah1944}, and stars from Ref.~\citenum{Grigoriev1996}.  Our calculated values of the specific heats appear as a solid line. The blue dashed and dotted vertical lines denote the Debye  and melting temperature respectively.}
\end{figure}

The optimized lattice parameters ${\bf x}_0(T)$ are found by fitting the free energy to a polynomial function~\cite{Tohei2016, Wang2006, Schmerler2014} of the lattice parameters $x_i=a, b$, and $c$:
\begin{equation}\label{sixth_order}
    F_{\textrm{fit}}({\bf x}) =\sum_{i,j} f_{i,j,k} a^i b^j c^k.
\end{equation}
$f_{i,j,k}$ are the coefficients of the polynomial fit. Here, we used fourth-order polynomials, and varied $a$ ($a$ and $c$) in the case of Cu (AlN).

To determine the minimum of Eq.~\eqref{sixth_order}, we use the constrained BFGS minimization method as implemented in the Scipy optimize package~\cite{Jones2001, Byrd1995, Zhu1997} within Python. After storing these minimizing parameters they are fit to an eighth-order polynomial as a function of temperature to account for the numerical noise in the free energy calculations. The CTE is then calculated by computing the derivative of the smoothed data~\cite{Slack1975} as
\begin{equation}\label{therm_expansion}
    \alpha_{x_i} = \frac{1}{x^0_i}\frac{d\ x_i}{dT} = \frac{d\ \ln(x_i)}{dT}
\end{equation}
for the different components of the CTE. Where $x^0_i$ is either the zero temperature lattice constant or the temperature-dependent lattice constant, as shown here. Using either of these produces practically identical results as assumed by Slack~\cite{Slack1975}.

With the integrated phonon density of states, the specific heat at constant volume can be calculated with \textsc{tdep} for the relaxed geometry. To determine the specific heat at constant pressure ($C_p$) the well-known thermodynamic relationship between the specific heat at constant volume, ($C_v$) and the trace of the thermal expansion coefficient ($\alpha$) calculated with Eq.~\eqref{therm_expansion}, can be used: 
\begin{equation}\label{cpcv}
C_p = C_v +VT\frac{\Tr (\alpha)^2}{\beta_T},
\end{equation}
where $\beta_T$ is the isothermal compressibility which is the inverse of the isothermal bulk modulus $B$. To determine $B$, we fit the free energy at a fixed temperature versus volume of the cell using the Birch-Murnaghan equation of state~\cite{Birch1947,Birch1978,Murnaghan1944}. By fitting this equation of state, we can extract the cohesive energy, isothermal bulk modulus, and the pressure derivative of the bulk modulus as a function of temperature. 

\section{Results}\label{sec:2}
To test the accuracy of our \textsc{vasp} and \textsc{tdep} calculations and the accuracy of our algorithm, we will compare our calculated values to experimental results for well-known systems: bulk Cu and AlN. Bulk Cu is a face-centered cubic structure (space group $Fm\bar{3}m$) from 0 K to its melting point of approximately 1357~K~\cite{Brand2006,Preston-Thomas1990}. Bulk AlN has a wurtzite crystal structure (space group $P6_3mc$) from 0 K to its melting point of approximately 3270 K~\cite{MacChesney1970}. For each of these materials, our calculations can be compared to experimental measurement as shown in Table~\ref{compare_values} and in the figures within the text. Table~\ref{compare_values} contains a comparison of our ground state (0~K) calculated Debye temperature, isentropic bulk modulus, and DFT lattice parameters to experimental measurement. The agreement between our values and experiment is well within the typical errors seen in ground state GGA-DFT calculations.
\begin{figure}[t!]
\includegraphics[width=0.4\textwidth]{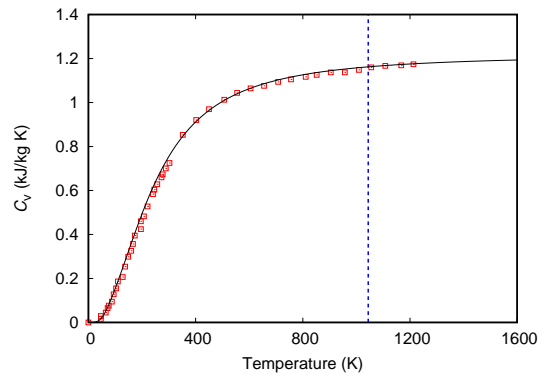}
\caption{\label{comp_cv} Calculated specific heat $C_v$ for AlN. Experimental data from Ref.~\citenum{Koshchenko1985} is given as points and our calculated values of the specific heat appear as a solid black line. The blue dashed vertical line denotes the Debye temperature.}
\end{figure}

\begin{table}[!t]
\caption{Our ground state (0~K) calculated Debye temperature, isentropic bulk modulus, and lattice parameters compared to experimental values.  The experimental lattice parameters of Cu and AlN come from experimental data extrapolated to zero temperature. }
\label{compare_values}
\centering
\begin{tabular}{|l|d{3.3} d{3.3} c |d{4.3} d{4.3} c |}\toprule
& \multicolumn{3}{c|}{Cu} & \multicolumn{3}{c|}{AlN} \\ \hline
& \multicolumn{1}{c}{ Calc.} &  \multicolumn{1}{c}{ Exp.} &  \multicolumn{1}{l|}{ Ref.} & \multicolumn{1}{c}{ Calc.} &  \multicolumn{1}{c}{ Exp.} &  \multicolumn{1}{l|}{ Ref.}  \\ \hline
$\Theta_D$ (K) & 348  &347 & \citenum{Stewart1998}& 1048 & 971 & \citenum{Wang2014} \\ \hline
$B$ (GPa) & 143.8  &137.6 & \citenum{Ledbetter2009} & 196.5 & 201 &\citenum{Wright1997} \\ \hline
$a$ (\AA)& 3.633 & 3.615 & \citenum{Giri1985} &3.129  & 3.107 & \citenum{Iwanaga2000} \\ \hline
$c$ (\AA)&   & &  & 5.012 & 4.973 & \citenum{Iwanaga2000}\\ \hline
  \botrule      
\end{tabular}
\end{table}

The temperature dependent lattice parameters of Cu and AlN are shown in Figs.~\ref{thermal_lattice_copper} and~\ref{thermal_lattice_aln}, respectively. The difference between our calculated values and experimental measurement comes from the inherent error in predicting the lattice parameters at the specific level of theory (i.e.: the PBE GGA of DFT). To compare to experiment over a broad temperature range for Cu, we have used the recommended values of the thermal lattice expansion, given in Ref.~\citenum{Touloukian1975}. 
\begin{figure}[b!]
\includegraphics[width=0.4\textwidth]{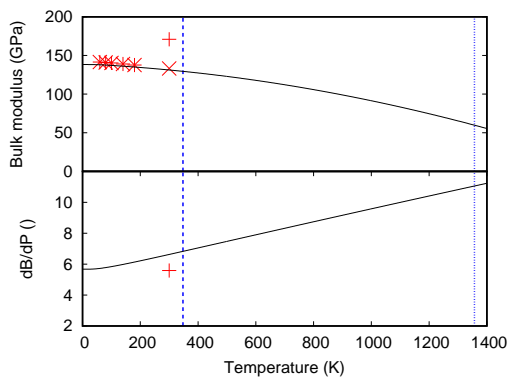}
\caption{\label{bulk_copper} Calculated isothermal bulk modulus, and its pressure derivative, for Cu. Upper panel: Experimental data points for the isothermal bulk modulus are shown in red: plus signs from Ref.~\citenum{Bridgman1931} for the isoentropic bulk modulus, crosses from Ref.~\citenum{Schmunk1960}, and asterisks from Ref.~\citenum{Shah1944} are for the isothermal bulk modulus. Lower panel: Experimental data for the unit-less pressure derivative at room temperature, shown as a red plus sign, comes from Ref.~\citenum{Bridgman1931} using the isoentropic bulk modulus.}
\end{figure}

We performed a polynomial fit to the extracted lattice parameters, shown as the solid lines in Figs.~\ref{thermal_lattice_copper} and~\ref{thermal_lattice_aln}. In addition we plot lattice parameters shifted by the ground-state error (difference between DFT calculated and experimentally extrapolated values at $T=0$~K) with black dashed lines for the sake of comparison. The un-shifted polynomial fit was used to calculate the CTE, specific heat, and bulk modulus. We obtain excellent agreement between the calculated CTE and experimental measurements for Cu and AlN between approximately 40 K and the Debye temperature, as shown in Figs.~\ref{thermal_expansion_copper} and \ref{thermal_expansion_aln}. However, there is some deviation above the Debye temperature, most notably for Cu where the calculated CTE is markedly higher than the experimental values as the temperature approaches the melting temperature. This is most likely due to higher-order anharmonic interactions not sufficiently accounted for at these temperatures. For AlN, it has been shown that the temperature dependent $c/a$ ratio is not correctly reproduced in anharmonic DFT/\textsc{tdep} calculations~\cite{Shulumba2016}, and our extracted anisotropic CTE of AlN are consistent with those results.  

The calculated specific heats of these materials also agree very well with experimental measurement as shown in Figs.~\ref{comp_cv_copper} and~\ref{comp_cv}. We also calculated $C_p$ for both materials, but the difference between $C_v$ and $C_p$ was too small over the entire temperature range of data to be discernible in the plot---at any temperature the difference $1-(C_p/C_v)$ was smaller than $10^{-6}$. This calculation, therefore, is further evidence that one can safely ignore the differences between these two quantities when comparing experimental and theoretical values~\cite{Cardona2007}.

\begin{figure}[b!]
\includegraphics[width=0.4\textwidth]{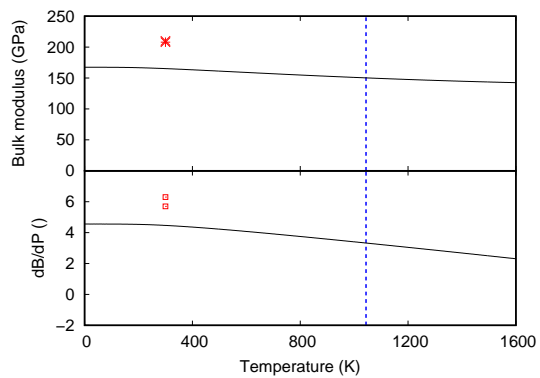}
\caption{\label{bulk_AlN} Calculated isothermal bulk modulus (upper panel), and its pressure derivative (lower panel), for AlN. Experimental data points for the isoentropic bulk modulus are given as red points: X from Ref.~\citenum{McNeil1993}, and plus sign from Ref.~\citenum{Ueno1992}.  Data for the unit-less pressure derivative comes from Ref.~\citenum{Xia1993}.}
\end{figure}

Finally, during our fitting of the equation of state, we extract the isothermal bulk modulus and its pressure derivative at zero external pressure as a function of temperature. The isothermal bulk modulus and the CTE are used to calculate the specific heat at constant pressure for both Cu and AlN.  In Figs.~\ref{bulk_copper} and~\ref{bulk_AlN}, we plot the isothermal bulk modulus and pressure derivative vs temperature for both Cu and AlN, respectively. Literature values of the isoentropic bulk modulus are shown for the sake of comparison and lie above our extracted values. A technical explanation of our method is outlined in the appendix.

\section{Discussion and Conclusion}\label{sec:3}
Our algorithm calculates the thermal properties of both isotropic and anisotropic materials using the DFT, DFPT and TDEP methods. Several related thermal properties are also calculated including the isothermal bulk modulus and specific heat at constant pressure. Agreement between our calculations and experimental results for the temperature dependent lattice parameters, specific heat at constant pressure, and isothermal bulk modulus give us confidence in our calculation methods. The correspondence between our calculations and experiment for the CTE is only moderate at temperatures above the Debye temperature.  This may e.g.\ be due to unattributed numerical inaccuracies, the present level of theory (PBE-GGA), or due to an inadequate sampling of the anharmonic interactions at elevated temperatures due to the finite number of configurations used in this calculation.  Nevertheless, most of our predictions are very well aligned with experiment, which makes us confident that this software can be used to determine the CTE for many more materials of experimental interest either with individual calculations or as part of a high-throughput framework.  

We have undertaken several careful convergence studies of the number of configurations, super-cell size, and $k$-point mesh density to minimize the error between our calculations and experimental measurement over the entire temperature range. We have determined that using the $k$-point mesh density given above provides a reasonable accuracy between our calculated CTE and experiment. Additionally, we have determined that the configuration temperature used by \textsc{tdep} should be slightly smaller than the Debye temperature to account for a reasonable amount of the anharmonic interactions as the temperature approaches the melting temperature.  Here, we show results with the configuration temperature equal to 80\% of the Debye temperature and have tested both higher and lower configuration temperatures as well with similar results.

While not shown here, the present algorithm can be used to calculate the CTE from a completely anisotropic system. Additionally, this algorithm can easily be integrated into existing high-throughput software workflows that we hope will enable subsequent researchers in e.g.\ designing the next generation temperature-dependent electronics. While one must use caution when calculating temperature-dependent properties, and carefully consider the convergence of many of the DFT, DFPT, and tdep input parameters, this software thus provides an algorithm for automatically calculating the CTE and related temperature-dependent properties of any material with a periodic unit cell. The symmetry and size of this unit cell may be limiting factors of the choice of materials, since it may be prohibitively expensive to perform calculations on very large unit cells with low symmetry with currently available high performance computing resources.

\section*{Acknowledgments}
We would like to thank the Research Council of Norway through the Frinatek program for funding and Martin Fleissner Sunding, Olle Hellman, and Nina Shulumba for helpful conversations and support. The computations were performed on resources provided by UNINETT Sigma2 - the National Infrastructure for High-Performance Computing and Data Storage in Norway through grant numbers nn2615k and nn9462k.

\appendix
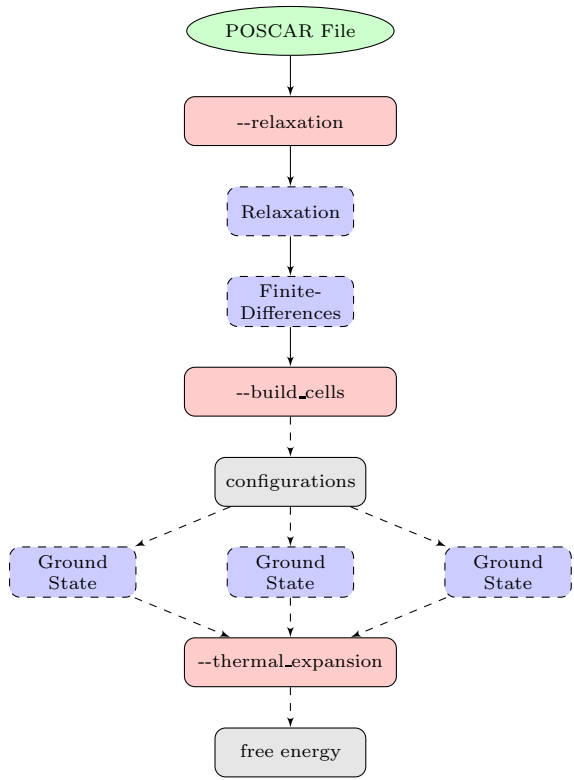
\begin{figure}[t!]
\begin{tikzpicture}[scale = 1,node distance = 1.2cm, auto,font=\scriptsize]
    \node [cloud] (input) {POSCAR File};
    \node [decision,below of = input] (bash1) {-{}-relaxation};
    \node [blockp,below of  = bash1] (rlx) {Relaxation};
    \node [blockp,below of =rlx](els) {Finite-Differences};
    \node [decision,below of = els] (bash2) {-{}-build\_cells};
    \node [blockt,below of = bash2] (tdep1) {configurations};
    \node [blockp,below of = tdep1] (free){Ground State};
    \node [blockp,left=1.2cm of free] (free1){Ground State};
    \node [blockp, right=1.2cm of free] (free2){Ground State};
    \node [decision,below of = free] (props) {-{}-thermal\_expansion};
    \node [blockt,below of = props] (tdep2) {free energy};

    \path [line,solid] (input)- -(bash1);
     \path [line,solid] (bash1)- -(rlx);
    \path [line,solid] (rlx) - - (els);
    \path [line,solid] (els) - - (bash2);
     \path [line,dashed] (bash2) - - (tdep1);
    \path [line,dashed] (tdep1) - - (free);
    \path [line,dashed] (tdep1) - - (free1);
    \path [line,dashed] (tdep1) - - (free2);
    \path [line,dashed] (free) - - (props);
    \path [line,dashed] (free1) - -(props);
    \path [line,dashed] (free2) - -(props);
    \path [line,dashed] (props) - - (tdep2);
 
\end{tikzpicture}
\caption{\label{fig:outline}Flow-chart outlining the algorithm used to calculate the anisotropic CTE.  Blue rectangles represent first-principles calculations using \textsc{vasp} and red rectangles represent calculations done with our script.  \textsc{tdep} calculations (grey rectangles) are automatically launched by our script during the last two stages of the calculation. Solid rectangles and lines correspond to serial calculations and dashed lines and rectangles represent parallel calculations.}
\end{figure}
\section{Summary of the Algorithm}\label{sec:1}
Our algorithm for calculating the CTE and associated thermal properties is diagrammed in Fig.~\ref{fig:outline} which involves both serial and parallel computations using \textsc{vasp} and \textsc{tdep}. To start a calculation, the user enters a single \textsc{vasp} formatted POSCAR file describing the material's crystal structure into the directory where the entire calculation will take place. This POSCAR file should represent the conventional cell of the system in the desired symmetry and, preferably, should represent the DFT-relaxed ground state structure. To launch the calculations, the user should execute the script with the tag "-{}-relaxation" which generates the necessary input files and bash scripts, creates the directories for the \textsc{vasp} calculations, and checks the convergence of the calculations once completed. Following a relaxation calculation to ensure convergence, the script will automatically launch the finite-differences calculations for the needed tensor quantities. Should the algorithm detect an error in the data or convergence procedure, the code will abort.  Once the error is fixed, the user can relaunch the calculation. 

After the finite-differences calculation, our algorithm automatically starts the next stage of the calculation (which can be manually executed with -{}-build\_cells) that constructs a set of unit cells corresponding to different volumes. The algorithm generates the necessary lattice parameter perturbations, as described above, to determine the CTE for the compound. During this calculation, the script will analyze the symmetry of the relaxed unit cell, calculate the Debye temperature for the creation of the initial set of configurations used by \textsc{tdep} to determine the inter-atomic force constants, and generate the necessary files for the related \textsc{vasp} calculations.  Once generated, the algorithm will rerun the relaxation of the newly perturbed unit cell by allowing relaxation of only the atomic positions. After this second relaxation, the script will generate the configurations and launch each configuration in parallel.  Since this is the most time-consuming part of the calculation, the user can relaunch this calculation as needed and only the parts of the calculation that are incomplete will be executed. 

The finishing stage of this calculation can be executed with the tag "-{}-thermal\_expansion" in which the script, for each lattice perturbation, will launch \textsc{tdep} to post-process the results of the previous \textsc{vasp} calculations. This post-processing generates the phonon density of states and various thermal properties including the free energy of the system. After gathering the calculated free energies, the script will determine the set of lattice parameters that minimize the free energy as a function of temperature. This temperature dependent set of lattice parameters is used to determine the CTE, isothermal bulk modulus, and specific heat at constant pressure as outlined above.

During the concluding stage of the calculation, several output files are produced. The calculated free energies for each volume as a function of temperature are printed to a single data file "out.free\_energy\_vs\_temp" with temperatures given in Kelvin and free energies given in eV. The temperature dependent lattice parameters are printed to a file called "out.thermal\_expansion" with the lattice parameters given in \AA. The thermal expansion coefficients, given in Eq.~\eqref{therm_expansion}, are printed to the file "out.expansion\_coeffs" in units of K$^{-1}$. Calculations of the isothermal bulk modulus, its pressure derivative, and cohesive energy are printed to the file "out.isothermal\_bulk" where the isothermal bulk modulus is given in GPa, the derivative of the bulk modulus with respect to pressure is unit-less, and the cohesive energy is in eV. Finally, the specific heat at constant volume and constant pressure are calculated using Eq.~\eqref{cpcv}.  These are printed to the data file "out.cv\_cp" where both specific heats are given in units of kJ kg$^{-1}$ K$^{-1}$. 

This software, and corresponding documentation, is available by direct request to the corresponding author.  

\bibliography{pyroelectric}

\end{document}